\documentclass[twocolumn,showpacs,preprintnumbers,amsmath,amssymb]{revtex4}

\usepackage{graphicx}
\usepackage{dcolumn}
\usepackage{bm}

\begin{document}

\preprint{APS/123-QED}

\title{Evidence for multiband superconductivity in the heavy fermion compound UNi$_2$Al$_3$}

\author{M. Jourdan}
 \email{jourdan@uni-mainz.de}
 \homepage{http://www.uni-mainz.de/FB/Physik/AG_Adrian/}
\author{A. Zakharov}
\author{M. Foerster}
\author{H. Adrian}
\affiliation{Institut f\"ur Physik, Johannes Gutenberg-Universit\"at, Staudingerweg 7, 55128 Mainz, Germany}

\date{\today}

\begin{abstract}
Epitaxial thin films of the heavy fermion superconductor UNi$_2$Al$_3$ with $T_c^{max}=0.98$K were investigated. The transition temperature $T_c$ depends on the current direction which can be related to superconducting gaps opening at different temperatures. Also the influence of the magnetic ordering at $T_N\simeq 5$K on $R(T)$ is strongly anisotropic indicating different coupling between the magnetic moments and itinerant charge carriers on the multi-sheeted Fermi surface . The upper critical field $H_{c2}(T)$ suggests an unconventional spin-singlet superconducting state.
\end{abstract}

\maketitle

\begin{figure}[htb]
\includegraphics[width=1.0\linewidth]{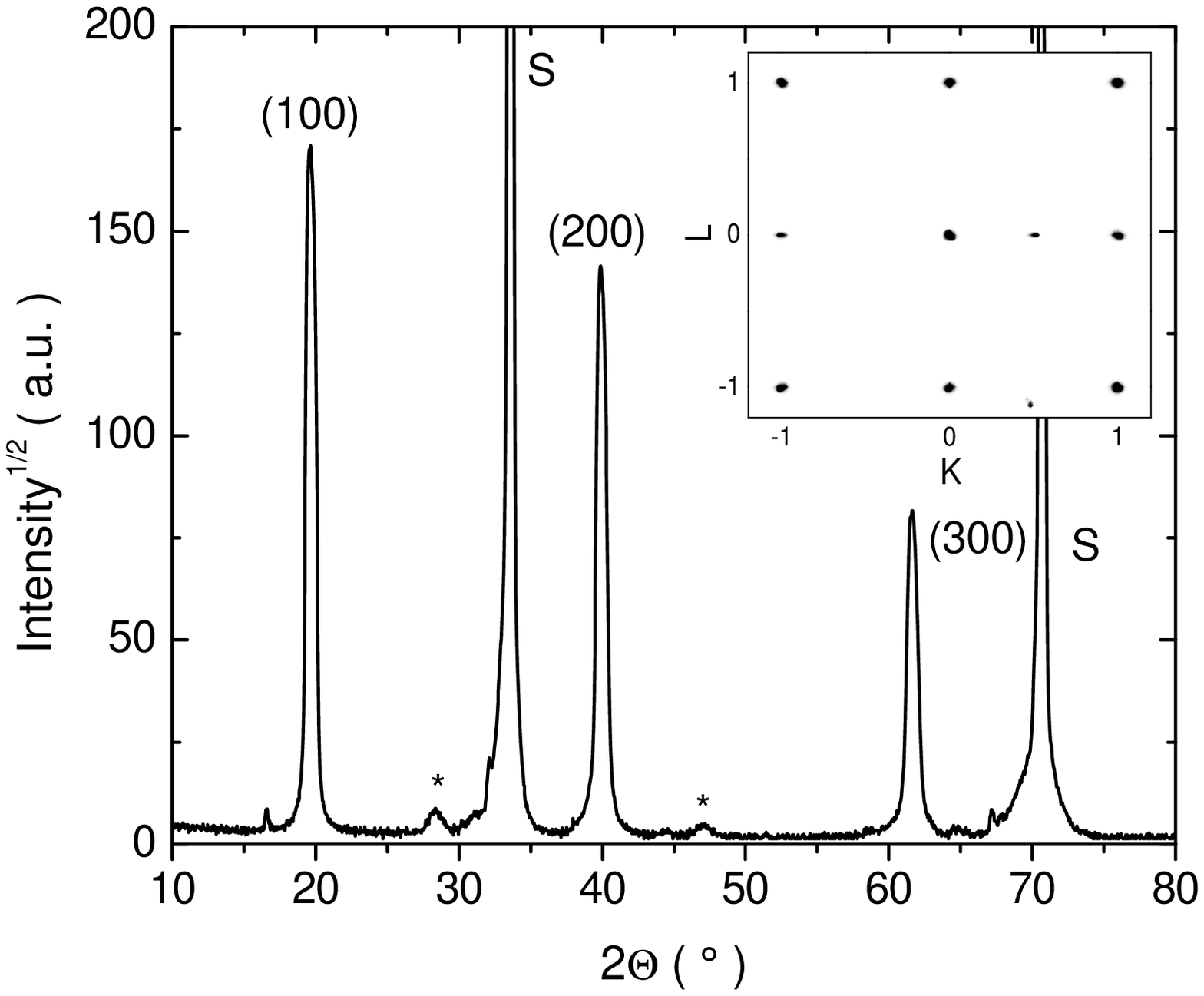}
\caption{\label{x-ray} X-ray $\Theta/2\Theta$-scan of an UNi$_2$Al$_3$ thin film on an YAlO$_3$(010) substrate. Please note the alignment of the reciprocal (100)-axis of the hexagonal crystal perpendicular to the substrate surface. (UNi$_2$Al$_3$ reflections are labeled (H00), substrate S, impurities *(presumably UO$_x$)). The inset shows a scan of the reciprocal (1KL)-plane of UNi$_2$Al$_3$. The x-ray intensity (linear grey scale) is plotted versus the momentum transfer in units of the reciprocal lattice vectors. From the observation of reflections at integer K and L values the in-plane order of the  film is obvious. The narrow reflections close to (1 1/2 -1) and (1 1/2 0) are substrate reflections.}
\end{figure}

\begin{figure}[htb]
\includegraphics[width=1.0\linewidth]{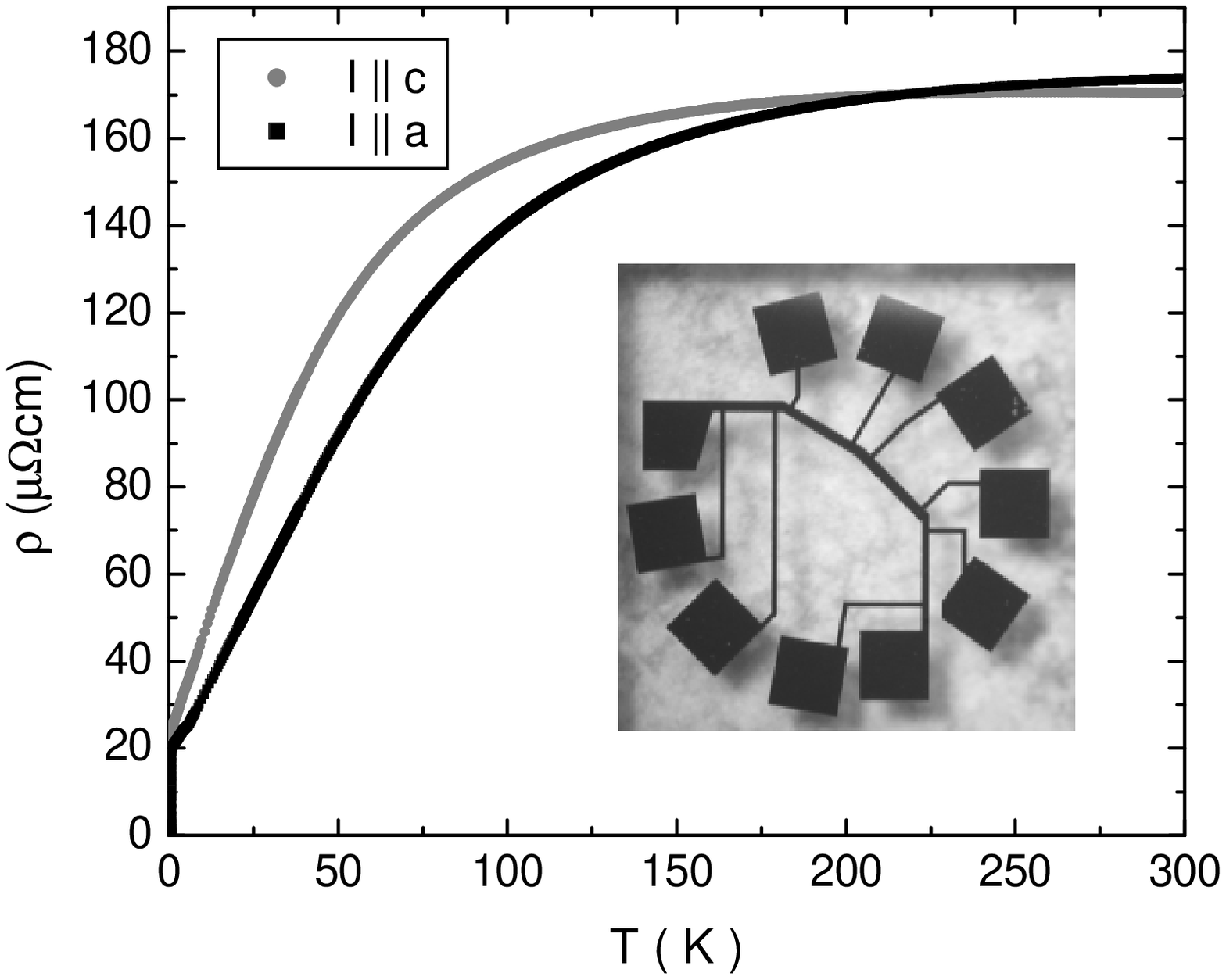}
\caption{\label{RT-all} Specific resistivity $\rho(T)$ of an UNi$_2$Al$_3$ thin film for different current directions (black squares: $I\parallel$ a, grey dots: $I\parallel$ c). Inset: Photograph of a patterned UNi$_2$Al$_3$ film ($4 \times 4$mm). For $R(T)$-measurements a current is send through the central conductor path (width d=100$\mu$m), the thin strips are voltage probes (4-point dc method).}
\end{figure}

\begin{figure}[htb]
\includegraphics[width=1.0\linewidth]{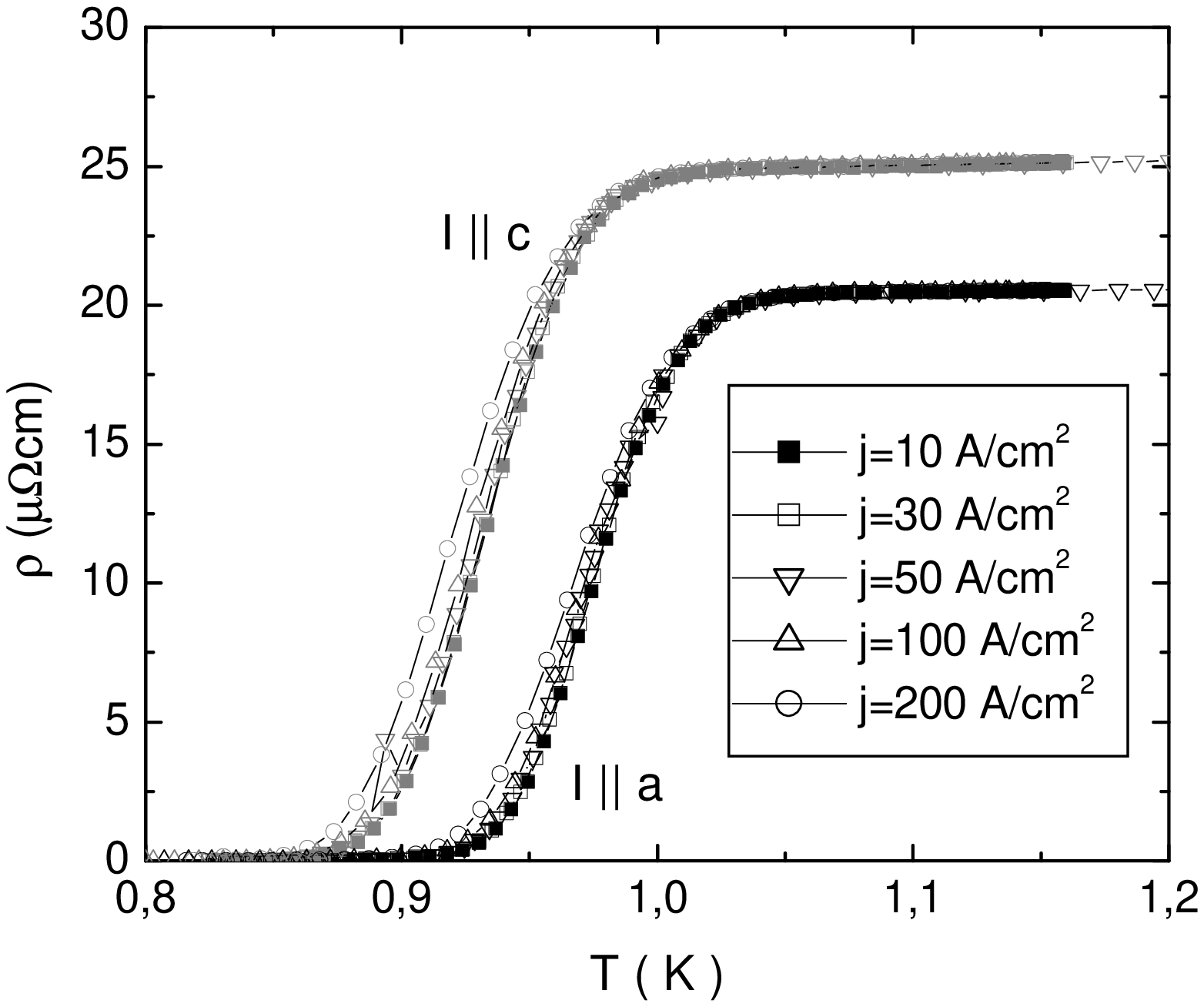}
\caption{\label{SC-trans} Superconducting resistive transitions of an UNi$_2$Al$_3$ thin film for different current directions (black symbols: $I\parallel$ a, grey symbols: $I\parallel$ c). Measurements with different current densities are show to demonstrate that the probe current does not influence the $T_c$ values (for $j\simeq 10$A/cm$^2$).}
\end{figure}

\begin{figure}[htb]
\includegraphics[width=1.0\linewidth]{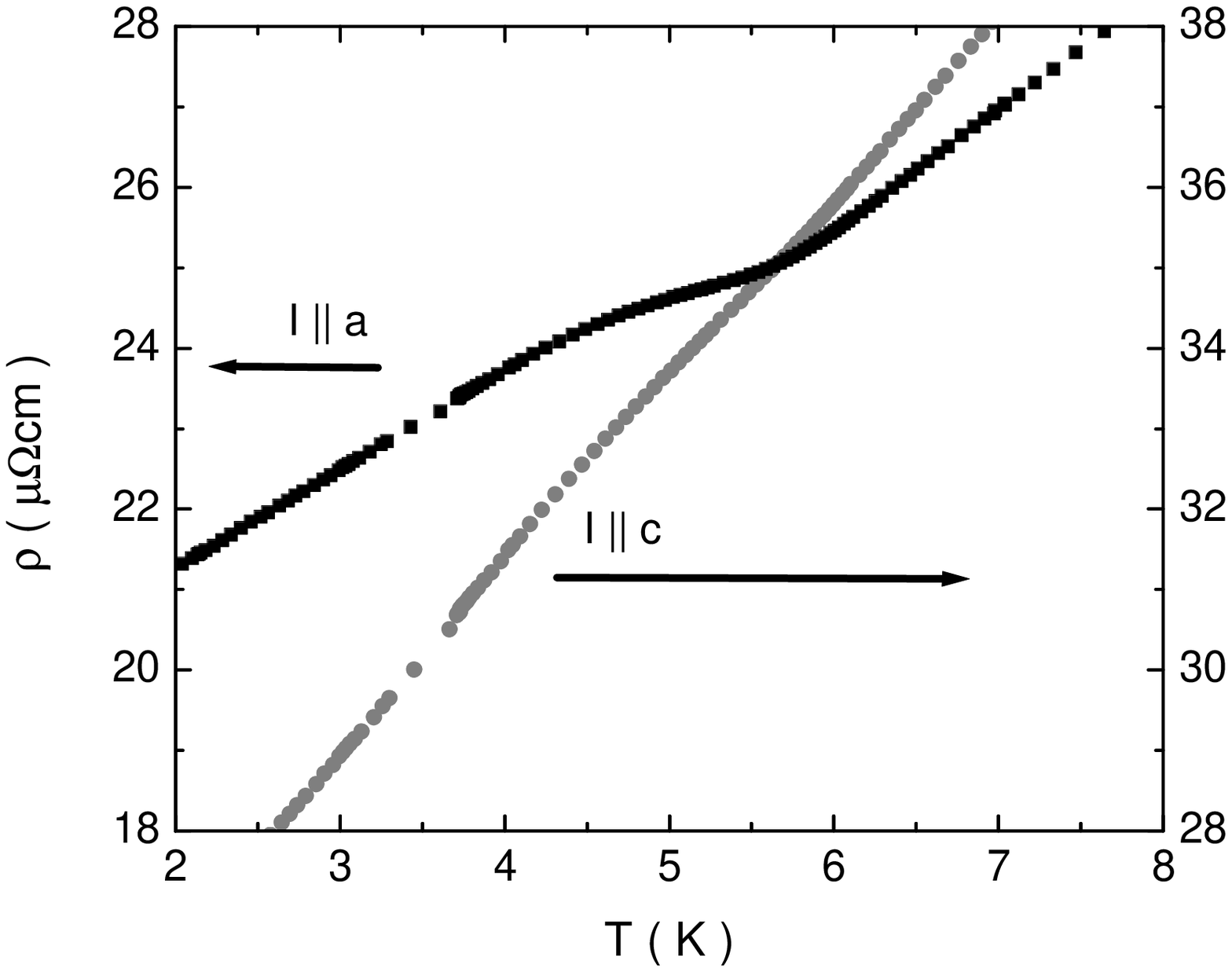}
\caption{\label{RT_TN} Specific resistivity $\rho(T)$ of an UNi$_2$Al$_3$ thin film. Left axis: $I\parallel$ a (black squares). Right axis: $I\parallel$ c (grey dots).}
\end{figure}

\begin{figure}[htb]
\includegraphics[width=1.0\linewidth]{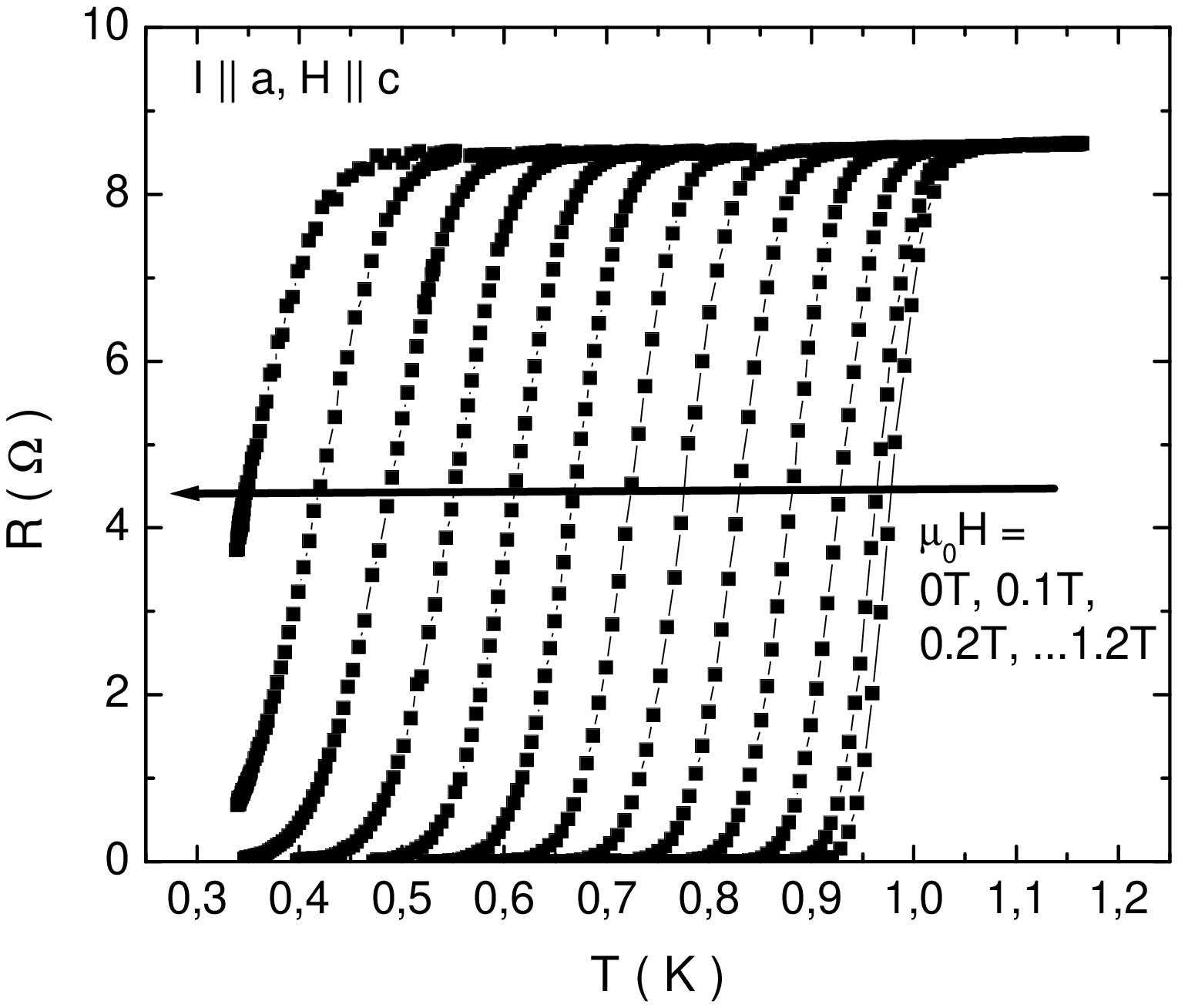}
\caption{\label{R(T)inB} Resistive superconducting transitions of an UNi$_2$Al$_3$ thin film in magnetic fields $\mu_0H$ from 0T to 1.2T. Probe current $I\parallel$ a and fields $H\parallel$ c.}
\end{figure}

\begin{figure}[htb]
\includegraphics[width=1.0\linewidth]{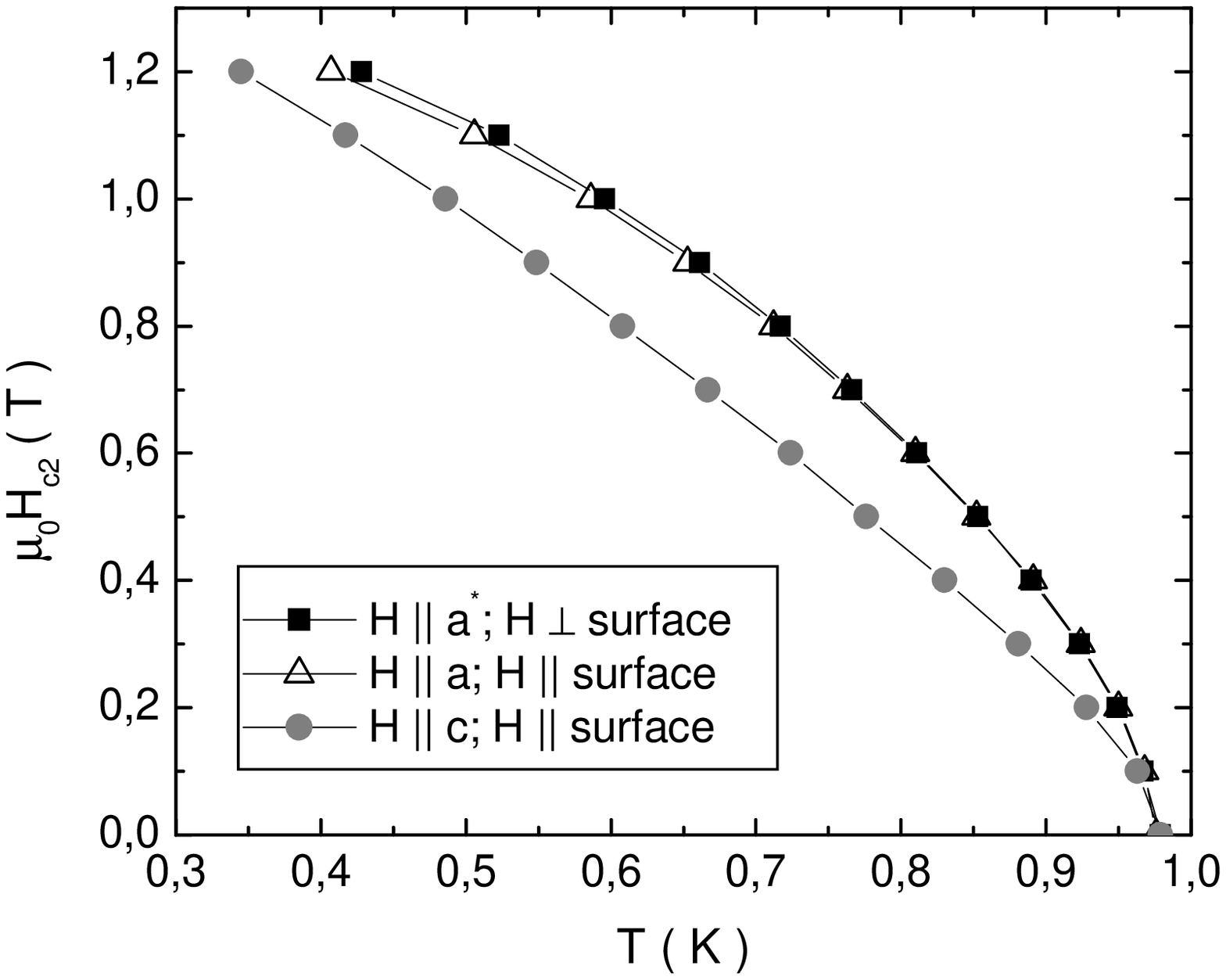}
\caption{\label{Bc2(T)} Upper critical fields $\mu_0H(T)$ of an UNi$_2$Al$_3$ thin film with different orientations relative to the field direction (see inset). Current $I\parallel$ a-axis.}
\end{figure}

Recently a clear progress concerning the superconductivity of the heavy fermion compound UPd$_2$Al$_3$ was reported. Due to combined experimental investigations based on tunneling spectroscopy on planar thin film junctions \cite{Jou99} and inelastic neutron scattering \cite{Met98, Ber98, Sat01} an unconventional mechanism of superconductivity was identified. There is compelling evidence that the Cooper pair formation in this compound is caused by the exchange of magnetic excitations. However, the question if these excitation are spin fluctuations \cite{Ber00} or  magnetic excitons \cite {Sat01,Tha02} is not finally answered. In this framework the investigation of the compound UNi$_2$Al$_3$, which is isostructural to UPd$_2$Al$_3$, is of great interest. Whereas UPd$_2$Al$_3$ shows a simple antiferromagnetic structure  with relatively large ordered magnetic moments of $\mu_{ord}\simeq0.85\mu_B$ the compound UNi$_2$Al$_3$ is an incommensurately ordered antiferromagnet \cite{Hie01} with smaller ordered moment $\mu_{ord}\simeq0.24\mu_B$ \cite{Sch94}. In contrast to UPd$_2$Al$_3$, for UNi$_2$Al$_3$ there is evidence for a spin-triplet superconducting state \cite{Ish02}.\\
The preparation of the first single crystalline bulk samples of UNi$_2$Al$_3$ \cite{Sat96,Mih97} was not possible until 5 years after the discovery of superconductivity in this compound \cite{Gei91}. This is due to a peritectical decomposition which result in the formation of an UAl$_2$ impurity phase \cite{Gei93} and hence a strongly reduced sample purity compared to UPd$_2$Al$_3$. However, since thin films are not prepared from the melt this typical problem can be avoided.
We prepared superconducting single crystalline thin films of UNi$_2$Al$_3$ (typical thickness $d\simeq 200$nm) by coevaporation of the elementary components in an MBE-system. Orthorhombical YAlO$_3$ cut in (010)- or (112)-direction provides an epitaxial substrate for the (100)-oriented UNi$_2$Al$_3$ with lattice misfits for the a-axis of +0.7\% for the former and -0.5\% for the latter substrate cut. On these substrates high purity single crystalline thin films were prepared. In Fig.\,\ref{x-ray} an x-ray $\Theta/2\Theta$-scan of an UNi$_2$Al$_3$ (100) film on YAlO$_3$(010) is shown. Only a very weak UO$_x$ but no UAl$_2$ impurity phase is visible. The inset shows a scan of the crystallographic (1KL)-plane of UNi$_2$Al$_3$. The observed x-ray reflections prove that the film is ordered in plane, i.\ e.\ single crystalline.    
From the positions of 14 x-ray reflections of an UNi$_2$Al$_3$ thin film the following lattice parameters were calculated: $a =0.5211(5)$nm, $b=0.5221(8)$nm, $c=0.4017(2)$nm. These numbers are in agreement with the values obtained from polycrystalline bulk samples \cite{Gei91} and rule out any growth process induced stress of the films.\\
Patterning the samples by a standard photolithographic process with ion beam etching made it possible to obtain well defined geometries for the investigation of anisotropic transport properties (see inset of Fig.\,\ref{RT-all}). Measurements of the temperature dependent resistivity $R(T)$ of the samples were performed with the current direction parallel to the crystallographic a-axis as well as parallel to the c-axis of the same thin film. A pronounced anisotropy, like reported previously from bulk single crystals \cite{Sul97} is observed (Fig.\,\ref{RT-all}). 
The residual resistivities of the best samples are $\rho_c=25\mu\Omega$cm for I $\parallel$ c and $\rho_a=20\mu\Omega$cm for I $\parallel$ a. The residual resistance ratios $RRR=R_{300K}/R_{1.1K}$ amount to $RRR_{I\parallel c}=6.6$ and $RRR_{I\parallel a}=8.3$. This is approximately half of the RRR reported for bulk single crystals \cite{Sat96, Sul97} indicating a higher concentration of scattering centers in the thin films. However, these scattering centers are less effective in destroying the superconducting state since the transition temperature T$_c$ and width $\Delta$T$_c$ of the films is as high and sharp as in the best bulk single crystals.\\
The films become superconducting at $T_c^{max}=0.98$K with resistive transition widths $\Delta T_c \simeq 0.06$K. Large critical current densities of $I_c \simeq 10^4$A/cm$^2$ at $T=0.32$K for $I\parallel$ a and $I\parallel$ c show that the observed superconductivity is a property of the complete thin film volume. The probe current employed for measurements of $T_c$ was adjusted to a current range were no influence of its magnitude on the resistive transition was observed (Fig.\,\ref{SC-trans}). All of our superconducting samples on both kinds of substrate show the superconducting transition for current direction $I \parallel $ c at a reduced temperature compared to $I \parallel$ a with $T_c(I\parallel$ a$)-T_c(I\parallel$ c$)\simeq 0.05$K (Fig.\,\ref{SC-trans}). The same behavior is visible at a close inspection of the only published $R(T)$-data of bulk single crystals by Sato et al.\ \cite{Sat96}. Thus there is strong evidence that the directional dependence of the resistive $T_c$ is an intrinsic property of UNi$_2$Al$_3$.\\
A possible explanation for this effect is the existence of two bands in UNi$_2$Al$_3$ with intraband pairing interactions but without superconducting interband interaction. Such a situation can result in two superconducting gaps on the Fermi surface which open at different temperatures \cite{Suh59}.
Up to now there is no Fermi surface calculation of UNi$_2$Al$_3$ available but similarities to the Fermi surface of the isostructural compound UPd$_2$Al$_3$ \cite{Kno96, Zwi03} can be assumed based on early band structure calculations \cite{Sti92}. Thus a multi-sheeted Fermi surface is expected with an increased anisotropy compared to UPd$_2$Al$_3$ as concluded from the stronger transport anisotropy of UNi$_2$Al$_3$. On such a Fermi surface superconducting gaps opening at different temperatures on different sheets could result in transitions temperatures which depend on the current direction.\\
According to ref.\,\cite{Suh59} a weak coupling between the two bands results in two gaps with a common $T_c$. However, in one of the bands an initially tiny energy gap is expected which opens drastically at a reduced temperature compared to $T_c$. In a transport experiment such a situation is presumably indistinguishable from the case of two different $T_c$ values due to critical current effects hiding the tiny energy gap. However, the simple model of ref.\,\cite{Suh59} presents only a first attempt to explain the directional dependence of $T_c$ and does not consider the coupling between superconductivity and magnetism  as well as Fermi surface and order parameter anisotropies of UNi$_2$Al$_3$.\\
At the magnetic ordering temperature $T_N\simeq5$K a clear anomaly in the resistivity $R(T)$ of the films is visible for currents $I\parallel$ a but not for $I\parallel$ c (Fig.\,\ref{RT_TN}).
The influence of the magnetic ordering on the resistivity of UNi$_2$Al$_3$ is different from the behavior of UPd$_2$Al$_3$ where a clear steepening of  $R(T)$ is observed on cooling beyond $T_N$ for $I\parallel$ a as well as for $I\parallel$ c \cite{Sat92}. In contrast, in the UNi$_2$Al$_3$ films we observe a flattening of $R(T)$ for $I\parallel a$ at $T_N$ which is similar to the behavior of URu$_2$Si$_2$ \cite{Map86}. This dependence can result from a change in the Fermi surface topology associated with the formation of a magnetization-density wave which opens a gap over a portion of the Fermi surface. Since $R(T)$ of UNi$_2$Al$_3$ is not affected by the magnetic ordering for currents $I\parallel c$ we conclude that only the sheet of the Fermi surface providing the a-axis transport couples to the magnetic order parameter.\\
The observation that the higher superconducting $T_c$ was observed on the Fermi surface sheet which is more affected by the magnetic ordering provides evidence that the Cooper-pairing in UNi$_2$Al$_3$ is mediated by magnetic excitations. However, due to the different influence of the magnetic ordering on the transport properties it is possible that the character of these excitations is different from the one discussed for UPd$_2$Al$_3$.\\ 
To investigate the superconducting order parameter concerning a possible spin triplet state the upper critical field $H_{c2}(T)$ of the UNi$_2$Al$_3$ thin films was determined by measuring resistive transitions $R(T)$ in different magnetic fields using a midpoint criterion. Fig.\,\ref{R(T)inB} shows the $R(T)$ curves obtained for probe currents $I\parallel$ a and field orientation $H\parallel$ c as an example.
Please note that the width and shape of the transitions depend only weakly on the field $H$ as opposed to the behavior reported from bulk single crystals \cite{Sat96}. The upper critical field was determined for field directions parallel to the real space a- and c-axis as well as parallel to the reciprocal a$^*$-axis of the hexagonal compound (Fig.\,\ref{Bc2(T)}). The measured $H_{c2}(T)$ curves are almost independent of the current direction. Only a small shift to lower temperatures of the data obtained with $I\parallel$ c compared to $I\parallel$ a was observed as expected considering the reduced $T_c$ in this direction.
For $H\parallel$ a$^*$ the film plane was perpendicular to the field direction, whereas for $H\parallel$ c the thin films had to be mounted with the film surface parallel to the field direction. In principle the parallel configuration allows the appearance of increased critical fields due to finite size effects (thin film and surface superconductivity) \cite{Tin96}. However, these phenomena should result in a pronounced angular dependence of $H_c(\Theta)$ which was not observed. For additional investigation of these possible effects the film was rotated by $90^{\rm o}$ on the sample platform 
resulting in a configuration with $H\parallel$ a parallel to the film surface. Due to the hexagonal symmetry of UNi$_2$Al$_3$ only a very weak 
anisotropy of the critical field is expected within the ab-plane. Thus the observation of very similar $H_{c2}(T)$-curves for $H\parallel$ a parallel to
the film surface and for $H\parallel$ a$^*$ perpendicular to the film surface shows that finite size effects do not influence the critical field.\\
For all investigated magnetic field directions a steep slope $-\mu_0H'_{c2}(T_c) > 5$T/K was observed in contrast to the much smaller slopes reported previously from investigations of bulk polycrystals \cite{Dal92} and single crystals \cite{Sat96}. Additionally, the critical fields of the thin films are much larger and less anisotropic than reported for the only available bulk single crystals \cite{Sat96}. Employing the conventional WHHM-theory \cite{Wer66} of the upper critical field the authors concluded that orbital pair breaking limits $H_{c2}$ of UNi$_2$Al$_3$. With the same approach \cite{Hak67} we obtain from
\begin {equation} \nonumber
H^{orb}_{c2}=0.693((-dH_{c2}/dT)_{T_c})T_c
\end {equation}  
an orbital upper critical field of $\mu_0H^{orb}_{c2}>3.5$T. However, from our $H_{c2}(T)$ curves we estimate $\mu_0H_{c2}(0$K$)\simeq 1.6$T only. This discrepancy provides evidence for a non-negligible paramagnetic pair breaking contribution and thus a spin singlet superconducting order parameter of UNi$_2$Al$_3$ as opposed to the experimental evidence for spin triplet superconductivity of ref.\,\cite{Ish02}.\\
Additional conclusions concerning the order parameter symmetry can not be drawn from the $B_{c2}(T)$ measurements. For a realistic description a concept which goes beyond the WHHM-theory considering the presumably multi-sheeted anisotropic Fermi surface and strongly anisotropic order parameter of UNi$_2$Al$_3$ is needed.\\
We arrive at the conclusion that epitaxial thin films of UNi$_2$Al$_3$ allow the identification of superconducting transition temperatures $T_c$ which clearly depend on the current direction. In conjunction with anisotropic signatures of the magnetic ordering these measurements provide evidence for a multiband superconducting state with a magnetic Cooper-pairing mechanism. Measurements of the upper critical field indicate a spin singlet superconducting state of UNi$_2$Al$_3$.\\
Financial support by the Materials Science Research Center (MWFZ) Mainz and the German Research Foundation (DFG) is acknowledged.

\end{document}